\documentclass[prl,twocolumn,amsmath,amssymb,superscriptaddress]{revtex4}
\usepackage{graphicx}

\newcommand{\mean}[1]{\left\langle #1 \right\rangle}
\newcommand{\unit}[1]{\,\mathrm{#1}}

\newcommand{\gam}{\gamma}
\newcommand{\m}{_\mathrm{m}}
\newcommand{\tot}{_\mathrm{tot}}
\newcommand{\tls}{two-level system}
\newcommand{\tm}{t_\mathrm{m}}

\begin{document}

\title{Measurement of Stochastic Entropy Production}

\author{C.~Tietz}
\affiliation{3. Physikalisches Institut, Universit\"at Stuttgart, 70550
  Stuttgart, Germany}
\author{S.~Schuler}
\affiliation{3. Physikalisches Institut, Universit\"at Stuttgart, 70550
  Stuttgart, Germany}
\author{T.~Speck}
\affiliation{{II.} Institut f\"ur Theoretische Physik, Universit\"at Stuttgart,
  70550 Stuttgart, Germany}
\author{U.~Seifert}
\affiliation{{II.} Institut f\"ur Theoretische Physik, Universit\"at Stuttgart,
  70550 Stuttgart, Germany}
\author{J.~Wrachtrup}
\affiliation{3. Physikalisches Institut, Universit\"at Stuttgart, 70550
  Stuttgart, Germany}

\begin{abstract}
  Using fluorescence spectroscopy we directly measure entropy production of a
  single two-level system realized experimentally as an optically driven
  defect center in diamond. We exploit a recent suggestion to define entropy
  on the level of a single stochastic trajectory (Seifert, Phys. Rev. Lett.
  {\bf 95}, 040602 (2005)).  Entropy production can then be split into one of
  the system itself and one of the surrounding medium. We demonstrate that the
  total entropy production obeys various exact relations for finite time
  trajectories.
\end{abstract}

\maketitle


Entropy as the central concept in statistical physics pervades many branches
of science. Well-defined and uncontested only in equilibrium, its extension to
time-dependent non-equilibrium phenomena has been debated since the days of
Boltzmann mostly in relation to an explanation of irreversibility and a
foundation of the second law of thermodynamics~\cite{lebo99a}. Major progress
arose with the formulation of the fluctuation theorem, which quantifies in the
long time limit the probability of entropy annihilating trajectories in small
systems constantly driven in a steady
state~\cite{evan93,evan94,gall95,kurc98,lebo99}. Entropy production in these
systems is either defined as phase space contraction rate or associated with a
dissipation functional, which ultimately should describe the dissipated heat.

By introducing the notion of a stochastic entropy along a single trajectory,
it has become possible both to extend the validity of the fluctuation theorem
to finite times and to prove an integral fluctuation theorem for the total
entropy production in arbitrarily driven systems governed by stochastic
dynamics~\cite{seif05a}. In this Letter, using our previously introduced
driven two-level system~\cite{schu05}, we measure the stochastic entropy
production along single trajectories and demonstrate that it obeys various
exact relations for finite times.

Fluctuation theorems for entropy production should be distinguished from
related theorems like the Jarzynski relation~\cite{jarz97}, Crooks'
theorem~\cite{croo99}, and the Hatano-Sasa relation~\cite{hata01}. The first
two allow to extract free energy differences from non-equilibrium work
measurements. Experimental tests have been performed by using a torsional
pendulum~\cite{doua05}, by mechanically stretching RNA
hairpins~\cite{liph02,coll05} as well as by driving a colloidal particle in a
time-dependent harmonic~\cite{wang02,carb04} and non-harmonic
potential~\cite{blic06}. The third one yields an exact relation for transition
between different steady states from which a general Clausius inequality
follows that has been tested using a driven colloidal particle~\cite{trep04}.
All these relations address a small system embedded in a surrounding heat bath
of constant temperature. In contrast, our set-up works athermally and does
therefore neither involve nor require any notion of dissipated heat. While our
previous work has demonstrated an exact Jarzynski-like relation for the
athermal analog of ``dissipated work''~\cite{schu05}, the present paper
addresses the concept of stochastic entropy production directly. The crucial
difference is that the derivation of the latter requires using the actual
nonequilibrium probability distribution in a general master
formula~\cite{seif05a}, whereas the former involves the corresponding
equilibrium distribution~\cite{seif04}. Still, both quantities fulfill various
exact relations like an integral fluctuation theorem.


\begin{figure*}
  \includegraphics[width=\linewidth]{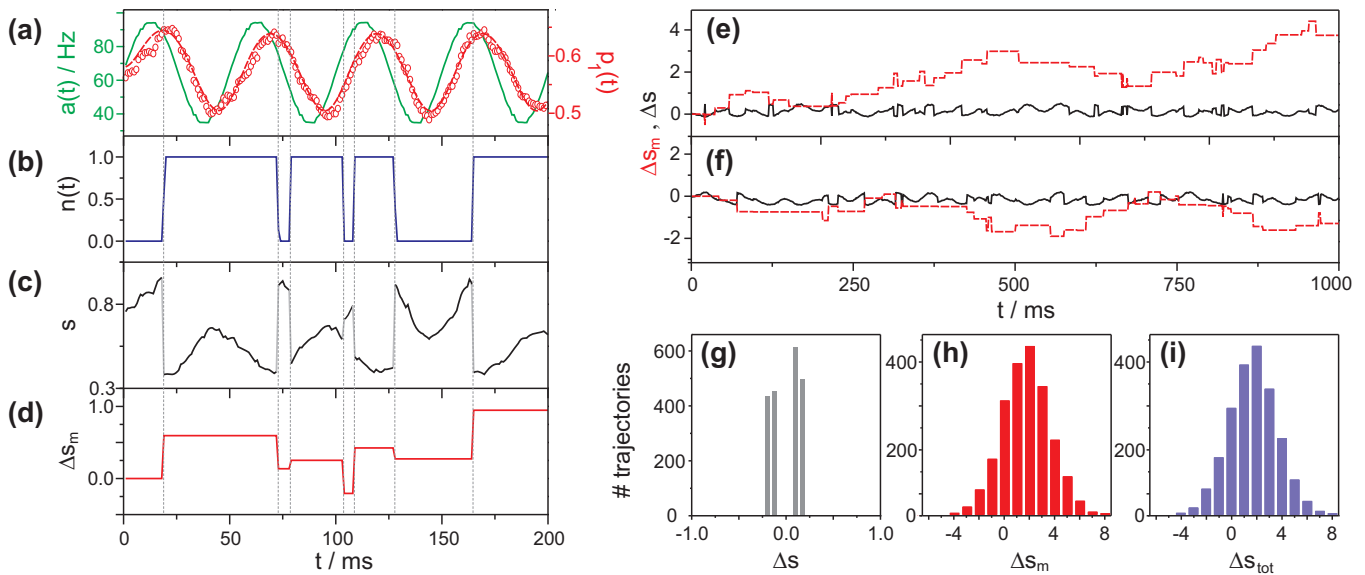}
  \caption{Entropy production in a single \tls\ with parameters
    $a_0=(15.6\unit{ms})^{-1}$, $b=(21.8\unit{ms})^{-1}$, $\tm=50\unit{ms}$,
    and $\gam=0.46$. (a) Transition rate $a(t)$ (solid/green line) and probability
    of the bright state $p_1(t)$ (red circles -- measured, dashed/red line --
    theoretical prediction) over 4 periods. (b) Single trajectory $n(t)$. (c)
    Evolution of the system entropy. The gray lines correspond to jumps
    (vertical dotted lines) of the system whereas the dark lines show the
    continuous evolution due to the driving. (d) Entropy change of the medium,
    where only jumps contribute. (e,f) Examples of (e) entropy producing and
    (f) entropy annihilating trajectories. The change of system entropy
    $\Delta s=s(t)-s(0)$ (solid/black) fluctuates around zero without effective
    entropy production, whereas in (e) $\Delta s\m$ (dashed/red) produces a net
    entropy over time.  In (f), $\Delta s\m$ consumes an entropy of about
    $1$ after 20 periods. (g-i) Histograms taken from 2,000 trajectories
    of the system (g), medium (h), and total entropy change (i). The system
    entropy shows four peaks corresponding to four possibilities for the
    trajectory to start and end ($0\mapsto1$, $1\mapsto0$, $0\mapsto0$, and
    $1\mapsto1$).  The distribution (h) of the medium entropy change has mean
    $\mean{\Delta s\m}=1.7$ and width $\sigma\m=3.7$, on this scale it
    differs only slightly from the distribution of the total entropy change
    (i).}
  \label{fig:1}
\end{figure*}

Our system is a photochromic defect center in natural IIa-type diamond. Its
optical properties indicate that we are dealing with a nickel-related
center~\cite{zait}. It can be excited by red light responding with a
Stokes-shifted fluorescence.  Additionally to this ``bright'' state the defect
exhibits a non-fluorescent ``dark'' state. The transition rates $a$ (from dark
to bright) and $b$ (from bright to dark) depend linearly on the intensity of
green and red light, respectively, turning the defect center into an effective
two-level system
\begin{equation}
  0\ (\text{dark}) \overset{a}{\underset{b}{\rightleftharpoons}} 1\ 
  (\text{bright})
\end{equation}
with controllable transition rates $a$ and $b$.

Single centers were addressed with a home-built confocal
microscope~\cite{grub97} using a dye laser (CR699, DCM) running at
$680\unit{nm}$ (red light intensity $\propto b$) superimposed to the
$514\unit{nm}$ line of an Ar-Ion laser (green light intensity $\propto a$) as
excitation sources.  While the red light was kept constant throughout the
experiment, the green light was modulated using a function generator
controlled acoustooptical modulator (AOM). In addition to the fluorescence of
the single defect, a second Avalanche Photo Diode recorded simultaneously the
alternating intensity of the green light.

The system can be found in state $n$ with probability $p_n$, where $n$ takes
either the value $0$ or $1$. To drive the system out of equilibrium, we
modulate the rate $a$ (from dark to bright) according to the sinusoidal
protocol
\begin{equation}
  a(t)=a_0[1+\gam\sin(2\pi t/\tm)]  
\end{equation}
whereas the rate $b$ is held constant. The parameters are the equilibrium
rates $a_0$ and $b$, the period $\tm$, and the modulation depth
$0\leqslant\gam<1$. At a time resolution of $1\unit{ms}$ the data of the two
detectors were acquired simultaneously after starting the modulation protocol
for the green laser. Following 20 periods with $\tm=50\unit{ms}$ and a certain
modulation depth $\gam$, the system was given $1,000\unit{ms}$ of unmodulated
green light to relax back into equilibrium. A sine function has been fitted to
the mean intensity of the green laser to obtain the modulation depth $\gam$.
The 4 runs in this Letter contain 3 times 1,000 and once 2,000 trajectories.
The modulation depths were varied from $\gam=0.07$ to $0.46$.



\begin{figure}
  \includegraphics[width=0.6\linewidth]{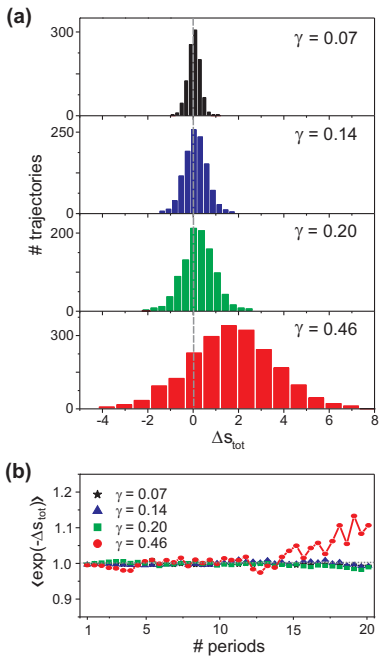}
  \caption{Experimental test of the integral fluctuation
    theorem~\eqref{eq:ift}. (a) Probability distribution of the total entropy
    production of a driven \tls. The modulation depth $\gam$ increases from
    $0.07$, $0.14$, $0.20$ to $0.46$; other parameters as in Fig.~\ref{fig:1}.
    The trajectory length is 20 periods with 1,000 trajectories per
    distribution (2,000 trajectories for $\gam=0.46$). (b) The mean
    $\mean{\exp[-\Delta s\tot]}$ over trajectory length for the four
    different modulation depths.}
  \label{fig:ift}
\end{figure}

A dimensionless, non-equilibrium entropy for driven systems on the level of a
single stochastic trajectory has been defined in Ref.~\cite{seif05a} as
\begin{equation}
  \label{eq:ent}
  s(t) = -\ln \left[ p_{n(t)}(t) \right],
\end{equation}
where the measured probability $p_n$ is evaluated at the actual state $n(t)$
at time $t$. Fig.~\ref{fig:1}a shows the protocol $a(t)$ together with the
probability $p_1(t)$ to dwell in the bright state. Fig.~\ref{fig:1}b displays
a sample binary trajectory $n(t)$ jumping between the two states. In
Fig.~\ref{fig:1}c we see that the evolution of $s(t)$ is governed by two
effects. First, the time-dependent driving of the rates leads to an evolving
probability resulting in a continuous contribution. Second, a jump between the
two states gives rise to a contribution $-\ln[p_+/p_-]$, where $p_-$ and
$p_+$ are the probabilities of the states immediately before and after the
jump, respectively.

Beside the entropy of the system itself, energy exchange and dissipation
lead, in general, to a change in medium entropy. For an athermal system, this
change in medium entropy $\Delta s\m$ can not be inferred from the exchanged
heat. Rather it has to be defined. In Ref.~\cite{seif05a}, the choice
\begin{equation}
  \Delta s\m = \ln\frac{w_{ij}}{w_{ji}}
\end{equation}
for a jump from state $i$ to state $j$ with instantaneous rate $w_{ij}$
($w_{ji}$ being the backward rate) has been motivated in analogy to the
thermal case. In our case it becomes $\Delta s\m=-\ln[a(t)/b]$ for a jump
$1\mapsto0$ and $\Delta s\m=-\ln[b/a(t)]$ for a jump $0\mapsto1$. As
demonstrated in Fig.~\ref{fig:1}d, the medium entropy changes only when the
system jumps, thereby balancing to some degree the change of $s(t)$.

One of the fundamental consequences of a stochastic entropy~\eqref{eq:ent} is
the fact that beside entropy producing trajectories also entropy annihilating
trajectories exist, see Fig.~\ref{fig:1}e and f, respectively. However, in
accordance with physical intuition, the latter become less likely for longer
trajectories or increased system size. Entropy annihilating trajectories not
only exist, they are essential to satisfy the integral fluctuation
theorem~\cite{seif05a}
\begin{equation}
  \label{eq:ift}
  \mean{\exp[-\Delta s\tot]} = 1.
\end{equation}
This theorem states that the average $\mean{\cdots}$ over infinitely many
realizations of a process involving the total entropy change $\Delta
s\tot=\Delta s+\Delta s\m$ becomes unity for any trajectory length and any
driving. Trajectories with $\Delta s\tot<0$ may occur seldom but are
exponentially weighted and thus contribute substantially to the left hand side
of Eq.~\eqref{eq:ift}. As an immediate consequence of Eq.~\eqref{eq:ift} one
has with $\mean{\Delta s\tot}\geqslant0$ a consistent formulation of the
second law of thermodynamic for small systems, giving {\it a posteriori}
support to the entropy definition~\eqref{eq:ent}.

Fig.~\ref{fig:ift}a demonstrates how the average entropy production increases
with increasing driving amplitude $\gam$. To fulfill the constraint imposed by
Eq.~\eqref{eq:ift} the distributions spread, making trajectories with large
negative production more likely. In Fig.~\ref{fig:ift}b we present the
experimental evidence for the validity of the theorem~\eqref{eq:ift} analyzing
1,000 trajectories. Only for the largest modulation depth in combination with
long trajectories ($t>15$ periods) a deviation from the theorem is observable.
This is due to the need for larger statistics as the mean value of the entropy
increases~\footnote{The discussion in Ref.~\cite{zuck02} with respect to the
  change of free energy can be extended to the present case.}.

\begin{figure}
  \includegraphics[width=0.8\linewidth]{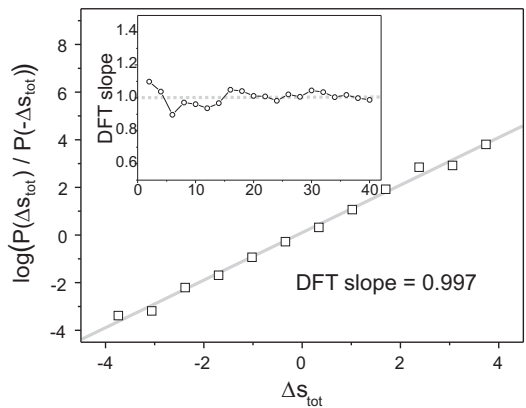}
  \caption{Experimental test of the detailed fluctuation theorem
    (DFT)~\eqref{eq:cft} after 20 periods with the same parameters as in
    Fig~\ref{fig:1}.  Following Eq.~\eqref{eq:cft}, the slope should be 1. The
    inset shows the slope of the DFT as function of the trajectory length.
    Due to the statistics of the entropy distribution, the DFT is satisfied
    best for longer trajectories.}
  \label{fig:cft}
\end{figure}
A stronger but also more special version of a fluctuation theorem is the
detailed fluctuation theorem~\cite{evan94,gall95}
\begin{equation}
  \label{eq:cft}
  \frac{P(+\Delta s\tot)}{P(-\Delta s\tot)} = \exp[+\Delta s\tot].
\end{equation}
Its experimental test is shown in Fig.~\ref{fig:cft}. Adapted to our
situation, it states that entropy annihilating trajectories with probability
$P(-\Delta s\tot)$ become exponentially less probable compared to trajectories
producing the same amount of entropy with the increasing absolute value of the
entropy. This theorem was derived originally for the long-time limit in
non-equilibrium steady states. However, following Crooks'
reasoning~\cite{croo99} it even holds for periodic driving as in our
experiment, provided (i) the protocol is time-symmetric, and (ii) the
distribution $p_n(t)$ has relaxed into the corresponding periodic limit
distribution. The first requirement is easily met by choosing start and end of
the trajectories at $n+1/4$ periods. As for the second requirement, in
Fig.~\ref{fig:1}a we see that the probability $p_n(t)$ indeed relaxes and
trails with a constant phase shift behind the driving rate $a(t)$. Therefore
we wait several periods before we start to record the data.


In conclusion, we have provided the first direct measurement of entropy
production along single stochastic trajectories in periodically driven
systems. In particular, we have shown that it is crucial to include the
entropy of the system itself. While for long trajectories it remains bounded,
its contribution is required for both the integral and the detailed
fluctuation theorem to hold for finite times. How this stochastic entropy will
contribute to a more comprehensive understanding of non-equilibrium dynamics
remains to be elucidated through the future study of more complex systems.

U.S. acknowledges support in part by the NSF under Grant No. PHY99-07949 while
this manuscript was finished.



\begin{thebibliography}{22}
\expandafter\ifx\csname natexlab\endcsname\relax\def\natexlab#1{#1}\fi
\expandafter\ifx\csname bibnamefont\endcsname\relax
  \def\bibnamefont#1{#1}\fi
\expandafter\ifx\csname bibfnamefont\endcsname\relax
  \def\bibfnamefont#1{#1}\fi
\expandafter\ifx\csname citenamefont\endcsname\relax
  \def\citenamefont#1{#1}\fi
\expandafter\ifx\csname url\endcsname\relax
  \def\url#1{\texttt{#1}}\fi
\expandafter\ifx\csname urlprefix\endcsname\relax\def\urlprefix{URL }\fi
\providecommand{\bibinfo}[2]{#2}
\providecommand{\eprint}[2][]{\url{#2}}

\bibitem[{\citenamefont{Lebowitz}(1999)}]{lebo99a}
\bibinfo{author}{\bibfnamefont{J.~L.} \bibnamefont{Lebowitz}},
  \bibinfo{journal}{Physica A} \textbf{\bibinfo{volume}{263}},
  \bibinfo{pages}{516} (\bibinfo{year}{1999}).

\bibitem[{\citenamefont{Evans et~al.}(1993)\citenamefont{Evans, Cohen, and
  Morriss}}]{evan93}
\bibinfo{author}{\bibfnamefont{D.~J.} \bibnamefont{Evans}},
  \bibinfo{author}{\bibfnamefont{E.~G.~D.} \bibnamefont{Cohen}},
  \bibnamefont{and} \bibinfo{author}{\bibfnamefont{G.~P.}
  \bibnamefont{Morriss}}, \bibinfo{journal}{Phys.\ Rev.\ Lett.}
  \textbf{\bibinfo{volume}{71}}, \bibinfo{pages}{2401} (\bibinfo{year}{1993}).

\bibitem[{\citenamefont{Evans and Searles}(1994)}]{evan94}
\bibinfo{author}{\bibfnamefont{D.~J.} \bibnamefont{Evans}} \bibnamefont{and}
  \bibinfo{author}{\bibfnamefont{D.~J.} \bibnamefont{Searles}},
  \bibinfo{journal}{Phys.\ Rev.\ E} \textbf{\bibinfo{volume}{50}},
  \bibinfo{pages}{1645} (\bibinfo{year}{1994}).

\bibitem[{\citenamefont{Gallavotti and Cohen}(1995)}]{gall95}
\bibinfo{author}{\bibfnamefont{G.}~\bibnamefont{Gallavotti}} \bibnamefont{and}
  \bibinfo{author}{\bibfnamefont{E.~G.~D.} \bibnamefont{Cohen}},
  \bibinfo{journal}{Phys.\ Rev.\ Lett.} \textbf{\bibinfo{volume}{74}},
  \bibinfo{pages}{2694} (\bibinfo{year}{1995}).

\bibitem[{\citenamefont{Kurchan}(1998)}]{kurc98}
\bibinfo{author}{\bibfnamefont{J.}~\bibnamefont{Kurchan}},
  \bibinfo{journal}{J.\ Phys.\ A:\ Math.\ Gen.} \textbf{\bibinfo{volume}{31}},
  \bibinfo{pages}{3719} (\bibinfo{year}{1998}).

\bibitem[{\citenamefont{Lebowitz and Spohn}(1999)}]{lebo99}
\bibinfo{author}{\bibfnamefont{J.~L.} \bibnamefont{Lebowitz}} \bibnamefont{and}
  \bibinfo{author}{\bibfnamefont{H.}~\bibnamefont{Spohn}},
  \bibinfo{journal}{J.\ Stat.\ Phys.} \textbf{\bibinfo{volume}{95}},
  \bibinfo{pages}{333} (\bibinfo{year}{1999}).

\bibitem[{\citenamefont{Seifert}(2005)}]{seif05a}
\bibinfo{author}{\bibfnamefont{U.}~\bibnamefont{Seifert}},
  \bibinfo{journal}{Phys.\ Rev.\ Lett.} \textbf{\bibinfo{volume}{95}},
  \bibinfo{pages}{040602} (\bibinfo{year}{2005}).

\bibitem[{\citenamefont{Schuler et~al.}(2005)\citenamefont{Schuler, Speck,
  Tietz, Wrachtrup, and Seifert}}]{schu05}
\bibinfo{author}{\bibfnamefont{S.}~\bibnamefont{Schuler}},
  \bibinfo{author}{\bibfnamefont{T.}~\bibnamefont{Speck}},
  \bibinfo{author}{\bibfnamefont{C.}~\bibnamefont{Tietz}},
  \bibinfo{author}{\bibfnamefont{J.}~\bibnamefont{Wrachtrup}},
  \bibnamefont{and} \bibinfo{author}{\bibfnamefont{U.}~\bibnamefont{Seifert}},
  \bibinfo{journal}{Phys.\ Rev.\ Lett.} \textbf{\bibinfo{volume}{94}},
  \bibinfo{pages}{180602} (\bibinfo{year}{2005}).

\bibitem[{\citenamefont{Jarzynski}(1997)}]{jarz97}
\bibinfo{author}{\bibfnamefont{C.}~\bibnamefont{Jarzynski}},
  \bibinfo{journal}{Phys.\ Rev.\ Lett.} \textbf{\bibinfo{volume}{78}},
  \bibinfo{pages}{2690} (\bibinfo{year}{1997}).

\bibitem[{\citenamefont{Crooks}(1999)}]{croo99}
\bibinfo{author}{\bibfnamefont{G.~E.} \bibnamefont{Crooks}},
  \bibinfo{journal}{Phys.\ Rev.\ E} \textbf{\bibinfo{volume}{60}},
  \bibinfo{pages}{2721} (\bibinfo{year}{1999}).

\bibitem[{\citenamefont{Hatano and Sasa}(2001)}]{hata01}
\bibinfo{author}{\bibfnamefont{T.}~\bibnamefont{Hatano}} \bibnamefont{and}
  \bibinfo{author}{\bibfnamefont{S.~I.}~\bibnamefont{Sasa}},
  \bibinfo{journal}{Phys.\ Rev.\ Lett.} \textbf{\bibinfo{volume}{86}},
  \bibinfo{pages}{3463} (\bibinfo{year}{2001}).

\bibitem[{\citenamefont{Douarche et~al.}(2005)\citenamefont{Douarche,
  Ciliberto, Petrosyan, and Rabbiosi}}]{doua05}
\bibinfo{author}{\bibfnamefont{F.}~\bibnamefont{Douarche}},
  \bibinfo{author}{\bibfnamefont{S.}~\bibnamefont{Ciliberto}},
  \bibinfo{author}{\bibfnamefont{A.}~\bibnamefont{Petrosyan}},
  \bibnamefont{and} \bibinfo{author}{\bibfnamefont{I.}~\bibnamefont{Rabbiosi}},
  \bibinfo{journal}{Europhys.\ Lett.} \textbf{\bibinfo{volume}{70}},
  \bibinfo{pages}{593} (\bibinfo{year}{2005}).

\bibitem[{\citenamefont{Liphardt et~al.}(2002)\citenamefont{Liphardt, Dumont,
  Smith, Tinoco~Jr, and Bustamante}}]{liph02}
\bibinfo{author}{\bibfnamefont{J.}~\bibnamefont{Liphardt}},
  \bibinfo{author}{\bibfnamefont{S.}~\bibnamefont{Dumont}},
  \bibinfo{author}{\bibfnamefont{S.~B.} \bibnamefont{Smith}},
  \bibinfo{author}{\bibfnamefont{I.}~\bibnamefont{Tinoco~Jr}},
  \bibnamefont{and}
  \bibinfo{author}{\bibfnamefont{C.}~\bibnamefont{Bustamante}},
  \bibinfo{journal}{Science} \textbf{\bibinfo{volume}{296}},
  \bibinfo{pages}{1832} (\bibinfo{year}{2002}).

\bibitem[{\citenamefont{Collin et~al.}(2005)\citenamefont{Collin, Ritort,
  Jarzynski, Smith, Tinoco, and Bustamante}}]{coll05}
\bibinfo{author}{\bibfnamefont{D.}~\bibnamefont{Collin}},
  \bibinfo{author}{\bibfnamefont{F.}~\bibnamefont{Ritort}},
  \bibinfo{author}{\bibfnamefont{C.}~\bibnamefont{Jarzynski}},
  \bibinfo{author}{\bibfnamefont{S.}~\bibnamefont{Smith}},
  \bibinfo{author}{\bibfnamefont{I.}~\bibnamefont{Tinoco}}, \bibnamefont{and}
  \bibinfo{author}{\bibfnamefont{C.}~\bibnamefont{Bustamante}},
  \bibinfo{journal}{Nature} \textbf{\bibinfo{volume}{437}},
  \bibinfo{pages}{231} (\bibinfo{year}{2005}).

\bibitem[{\citenamefont{Wang et~al.}(2002)\citenamefont{Wang, Sevick, Mittag,
  Searles, and Evans}}]{wang02}
\bibinfo{author}{\bibfnamefont{G.~M.} \bibnamefont{Wang}},
  \bibinfo{author}{\bibfnamefont{E.~M.} \bibnamefont{Sevick}},
  \bibinfo{author}{\bibfnamefont{E.}~\bibnamefont{Mittag}},
  \bibinfo{author}{\bibfnamefont{D.~J.} \bibnamefont{Searles}},
  \bibnamefont{and} \bibinfo{author}{\bibfnamefont{D.~J.} \bibnamefont{Evans}},
  \bibinfo{journal}{Phys.\ Rev.\ Lett.} \textbf{\bibinfo{volume}{89}},
  \bibinfo{pages}{050601} (\bibinfo{year}{2002}).

\bibitem[{\citenamefont{Carberry et~al.}(2004)\citenamefont{Carberry, Reid,
  Wang, Sevick, Searles, and Evans}}]{carb04}
\bibinfo{author}{\bibfnamefont{D.~M.} \bibnamefont{Carberry}},
  \bibinfo{author}{\bibfnamefont{J.~C.} \bibnamefont{Reid}},
  \bibinfo{author}{\bibfnamefont{G.~M.} \bibnamefont{Wang}},
  \bibinfo{author}{\bibfnamefont{E.~M.} \bibnamefont{Sevick}},
  \bibinfo{author}{\bibfnamefont{D.~J.} \bibnamefont{Searles}},
  \bibnamefont{and} \bibinfo{author}{\bibfnamefont{D.~J.} \bibnamefont{Evans}},
  \bibinfo{journal}{Phys.\ Rev.\ Lett.} \textbf{\bibinfo{volume}{92}},
  \bibinfo{pages}{140601} (\bibinfo{year}{2004}).

\bibitem[{\citenamefont{Blickle et~al.}(2006)\citenamefont{Blickle, Speck,
  Helden, Seifert, and Bechinger}}]{blic06}
\bibinfo{author}{\bibfnamefont{V.}~\bibnamefont{Blickle}},
  \bibinfo{author}{\bibfnamefont{T.}~\bibnamefont{Speck}},
  \bibinfo{author}{\bibfnamefont{L.}~\bibnamefont{Helden}},
  \bibinfo{author}{\bibfnamefont{U.}~\bibnamefont{Seifert}}, \bibnamefont{and}
  \bibinfo{author}{\bibfnamefont{C.}~\bibnamefont{Bechinger}},
  \bibinfo{journal}{Phys.\ Rev.\ Lett.} \textbf{\bibinfo{volume}{96}},
  \bibinfo{pages}{070603} (\bibinfo{year}{2006}).

\bibitem[{\citenamefont{Trepagnier et~al.}(2004)\citenamefont{Trepagnier,
  Jarzynski, Ritort, Crooks, Bustamante, and Liphardt}}]{trep04}
\bibinfo{author}{\bibfnamefont{E.~H.} \bibnamefont{Trepagnier}},
  \bibinfo{author}{\bibfnamefont{C.}~\bibnamefont{Jarzynski}},
  \bibinfo{author}{\bibfnamefont{F.}~\bibnamefont{Ritort}},
  \bibinfo{author}{\bibfnamefont{G.~E.} \bibnamefont{Crooks}},
  \bibinfo{author}{\bibfnamefont{C.~J.} \bibnamefont{Bustamante}},
  \bibnamefont{and} \bibinfo{author}{\bibfnamefont{J.}~\bibnamefont{Liphardt}},
  \bibinfo{journal}{Proc.\ Natl.\ Acad.\ Sci.\ U.S.A.}
  \textbf{\bibinfo{volume}{101}}, \bibinfo{pages}{15038}
  (\bibinfo{year}{2004}).

\bibitem[{\citenamefont{Seifert}(2004)}]{seif04}
\bibinfo{author}{\bibfnamefont{U.}~\bibnamefont{Seifert}},
  \bibinfo{journal}{J.\ Phys.\ A:\ Math.\ Gen.} \textbf{\bibinfo{volume}{37}},
  \bibinfo{pages}{L517} (\bibinfo{year}{2004}).

\bibitem[{\citenamefont{Zaitsev}(2001)}]{zait}
\bibinfo{author}{\bibfnamefont{A.~M.} \bibnamefont{Zaitsev}},
  \emph{\bibinfo{title}{Optical Properties of Diamond}}
  (\bibinfo{publisher}{Springer-Verlag}, \bibinfo{address}{Berlin Heidelberg
  New York}, \bibinfo{year}{2001}).

\bibitem[{\citenamefont{Gruber et~al.}(1997)\citenamefont{Gruber,
  Dr{\"a}benstedt, Tietz, Fleury, Wrachtrup, and von Borczyskowski}}]{grub97}
\bibinfo{author}{\bibfnamefont{A.}~\bibnamefont{Gruber}},
  \bibinfo{author}{\bibfnamefont{A.}~\bibnamefont{Dr{\"a}benstedt}},
  \bibinfo{author}{\bibfnamefont{C.}~\bibnamefont{Tietz}},
  \bibinfo{author}{\bibfnamefont{L.}~\bibnamefont{Fleury}},
  \bibinfo{author}{\bibfnamefont{J.}~\bibnamefont{Wrachtrup}},
  \bibnamefont{and} \bibinfo{author}{\bibfnamefont{C.}~\bibnamefont{von
  Borczyskowski}}, \bibinfo{journal}{Science} \textbf{\bibinfo{volume}{276}},
  \bibinfo{pages}{2012} (\bibinfo{year}{1997}).

\bibitem[{\citenamefont{Zuckerman and Woolf}(2002)}]{zuck02}
\bibinfo{author}{\bibfnamefont{D.~M.} \bibnamefont{Zuckerman}}
  \bibnamefont{and} \bibinfo{author}{\bibfnamefont{T.~B.} \bibnamefont{Woolf}},
  \bibinfo{journal}{Phys.\ Rev.\ Lett.} \textbf{\bibinfo{volume}{89}},
  \bibinfo{pages}{180602} (\bibinfo{year}{2002}).

\end{thebibliography}
\end{document}